# Dielectric properties of Poly(methyl methacrylate) (PMMA)/$CaCu_3Ti_4O_{12}$ Composites

P. Thomas,[a*] R.S. Ernest Ravindran,[a] K.B.R. Varma.[b]

[a] Dielectric Materials Division, Central Power Research Institute, Bangalore : 560 080, India
[b] Materials Research Centre, Indian Institute of Science, Bangalore: 560012, India
E-mail:*thomas@cpri.in

**Abstract:** Materials with high dielectric constant are in great demand for the miniaturization of electronic devices. More specifically, high dielectric constant polymer-ceramic composites are useful for embedded capacitor applications. A composite consisting of giant dielectric $CaCu_3Ti_4O_{12}$ (CCTO) incorporated into the Poly(methyl methacrylate) (PMMA) polymer matrix has been fabricated by melt mixing followed by hot pressing. The composites thus fabricated has been characterised for structural, morphological and dielectric properties. The composites dielectric constants had increased when the CCTO content increased in the PMMA matrix. The dielectric constant of PMMA is around 4.9 @ 100Hz which has increased to 15.7 @ 100Hz when the ceramic content has increased to 40 Vol %. At low frequencies, space charge polarisation is dominant. This composite also exhibited remarkably low dielectric loss at high frequency, which makes this composite a suitable candidate for the capacitors in high frequency application.

Kew words: Dielectric, Polymer composite, PMMA, $CaCu_3Ti_4O_{12}$.

## INTRODUCTION

Materials that exhibit high dielectric constant have become promising materials particularly for capacitors and memory devices. The $CaCu_3Ti_4O_{12}$ (CCTO) ceramic that exhibited giant permittivity has been used as filler for the fabrication of high dielectric constant composites for potential capacitor applications and studied in great detail [1-9]. It is generally observed that the dielectric constant increases as the CCTO content increases in the polymer and decreases as the frequency increases [1-9]. The Poly (Methyl Methacrylate) (PMMA), a transparent thermoplastic polymer, possess moderate properties and low cost and PMMA-ceramic composites were studied in great detail [10,11]. Though the dielectric constant obtained for PMMA-ceramic composites [10-12] are not so high compared to PVDF-CCTO composites [1-9], the dielectric loss obtained for PMMA based composites are significantly lower. It is interesting to note that the dielectric constant obtained for poly(methyl methacrylate)/multi-walled carbon nanotube is independent of frequency [13]. The recent work on the PMMA/MMT/CCTO ternary composites also exhibited improved dielectric properties [14]. Hence, we thought it is worth investigating the PMMA/CCTO composites as a suitable dielectric material for capacitors, as CCTO exhibit high dielectric constant [15], which is nearly independent of frequency (upto 10 MHz) and temperature (room temperature to 300°C). In this work, the details pertaining to the fabrication and characterization of PMMA/$CaCu_3Ti_4O_{12}$ composites involving CCTO as filler and PMMA as matrix material are reported.

## EXPERIMENTAL

### Fabrication and Characterization

The solid state reaction route was adopted for synthesising CCTO ceramic [8,15]. The stoichiometric amounts of AR grade $CaCO_3$, CuO and $TiO_2$ was weighed, mixed using acetone and ball milled (300rpm) for 5h. The homogeneous mixture thus obtained has been dried in oven for about 1h. This stoichiometric mixture was taken in a re-crystallized alumina crucible and heated at 1000°C for 10h to obtain phase pure CCTO [8]. In order to get submicron particles, the CCTO powders were ball milled for about 12h using a planetary mill. PMMA, having Molecular weight of 1,10,000, Make: LG Corporation was used as matrix material. For the fabrication of composites, initially, the as received PMMA granules were heated at 210°C in Brabender Plasticorder (Model:PLE331) till the PMMA granules are thoroughly melted. To this melt, the CCTO powders were slowly added and mixed for 15 min at this temperature. The mixture was taken out from the Plasticorder and hot-pressed at this temperature to obtain a sheet of 100 $mm^2$ with 1.0 mm in thickness. The composites with varying concentrations (0 to 40 %) of CCTO by Volume were fabricated.

To examine the phase formation and the structure, an XPERT-PRO Diffractometer (Philips, Netherlands) was used. Scanning electron microscope (FEI-Technai TEM-Sirion) was used for the microstructure analysis of the composite samples. Since the surfaces of the as prepared composites are not smooth enough for carrying out the dielectric measurements, the composite sheets were polished using an emery paper to achieve smooth and parallel surfaces. Further, these samples were cleaned under ultra-sonication and subsequently the surfaces were electrode with silver paste and cured at 50°C. An LCR meter (Model: HP4194A) was used for the capacitance measurement as a function of frequency (100Hz–1MHz) in room temperature. The measurement accuracy of the instrument is less than 5%. The dielectric constant was calculated using the relation, $\varepsilon_r = C \times d / \varepsilon_o A$, where $C$ =capacitance, $d$ is the thickness of the sample, $\varepsilon_o = 8.854 \times 10^{-12}$ F/m and A is the area of the sample.

## RESULTS AND DISCUSSION

The X-ray diffraction patterns were recorded on the CCTO ceramic, as received PMMA and on the composites that were fabricated. It has been observed that the as received PMMA exhibited amorphous nature (Fig.1(a)). The as prepared CCTO powders (Fig 1(b)) is compared well with the ICDD data (01-075-1149) (Fig 1(c)) and with the pattern reported earlier [8]. The PMMA+CCTO-10 Vol % CCTO composite exhibited composite nature, whereas, the PMMA+CCTO-40 vol % CCTO composite, only CCTO peaks are dominant. This is due to the fact that the CCTO crystallites are dominant in the composite.

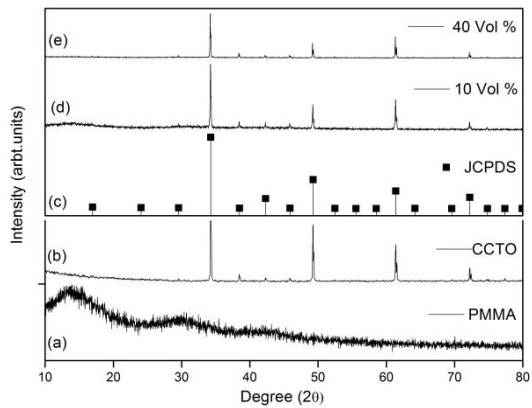

Fig.1. XRD patterns were recorded for (a) pure PMMA, (b) Phase-pure CCTO, (c) JCPDS data of CCTO, (d) PMMA+CCTO-10 Vol % and (e) PMMA+CCTO-40 Vol %.

Fig.2(a-d) gives the SEM micrographs recorded for the as prepared composites containing 2.8, 10, 21 & 40 Vol % of CCTO. It is seen that (Fig.2(b)) the crystallites are distributed evenly with less agglomeration associated with minimum porosity for PMMA+CCTO-10 Vol % composite. In the case of PMMA+CCTO-40 Vol % composite (Fig.2(d)), the crystallites are well packed and there is a better connectivity between the crystallites. The porosity is almost negligible for PMMA+CCTO-40 Vol % composite. It is reported that the even distribution of ceramic fillers in the polymer matrix and better connectivity between the crystallites is likely to enhance the dielectric constant of the composites [16]. We could achieve dielectric constant as high as 15.7 @ 100 Hz for the PMMA+CCTO-40 Vol % composites.

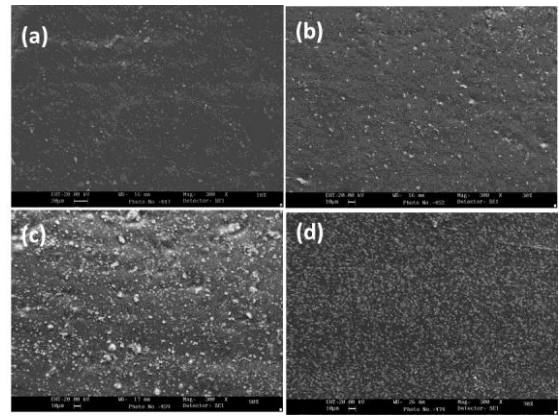

Fig.2. Scanning electron micrographs of (a) PMMA+CCTO-2.8 Vol %, (b) PMMA+CCTO-10 Vol%, (c) PMMA+CCTO-21 Vol% and (d) PMMA+CCTO-40 Vol % .

Fig.3 shows the room temperature dielectric constant ($\varepsilon_r^{'}$) and the dielectric loss (D) recorded in the frequency (100Hz-1MHz) for the CCTO ceramic derived from the solid state reaction route. The dielectric constant recorded @ 10kHz is 2500 and the loss is 0.04 respectively.

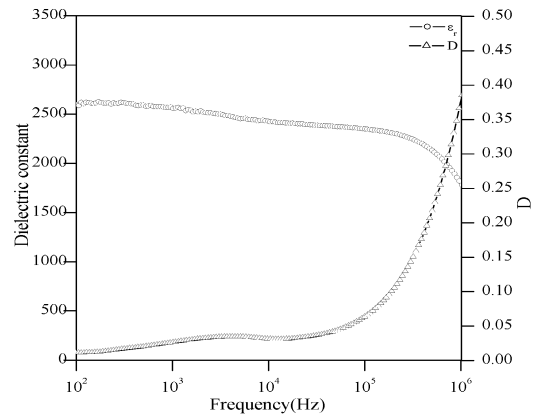

Fig.3.The dielectric constant and dielectric loss (D) as a function of frequency recorded at the room temperature for the CCTO ceramic sintered at 1100°C/2h

Fig.4. shows dielectric constant ($\varepsilon_r^{'}$) recorded for PMMA, and for the PMMA/CCTO composites at room temperature. It is seen that the dielectric constant of the PMMA is almost constant over the frequency under investigation. As expected, when the ceramic loading increases in the PMMA matrix, the dielectric constant increased at all the frequencies under study. The obtained dielectric constant for pure PMMA is around 4.9 @100Hz and it has increased to 15.7 @100Hz when the CCTO ceramic loading is increased to 40 Vol % in PMMA. The value of dielectric constant obtained for the PMMA+CCTO-40 Vol % is higher than the pure PMMA and much lower than that of CCTO (Fig.3). The similar results were reported, wherein, the introduction of $Al_2O_3$ ceramic into the PMMA matrix to the level of 45 vol %, increased the dielectric constant and it has been attributed to the space charge polarization mechanism [10]. In the case of PMMA+ $BaTiO_3$-40% composite that was fabricated by employing nanoceramics, dielectric constant obtained is only 19.5 [11]. However, the dielectric constant obtained for PMMA+$BaTiO_3$-40% is around 15.0 [12]. From these results, the marginal difference in the dielectric constants were observed are directly attributed to the particle size, demonstrating the particle size influence on the dielectric behaviour. In this work, CCTO with the particle size of around 1-7 μm has been used for the fabrication of composite. The PMMA+CCTO-40 Vol % composite (Fig.4) exhibited dielectric constant of 13.7 @1kHz, which has decreased to 11.7 @ 100MHz. However, frequency independent dielectric behaviour was noticed in the range 1kHz to 100MHz. At low frequency, space charge polarisation is dominant.

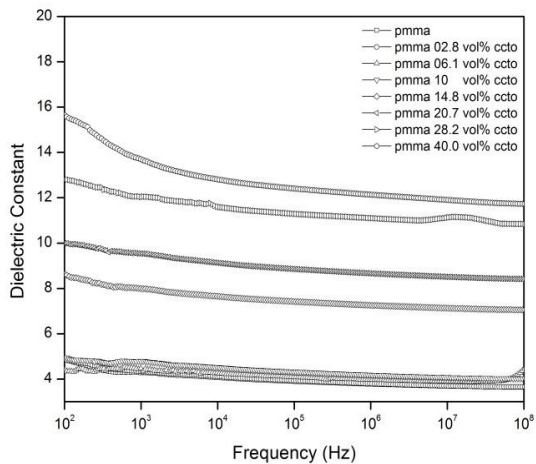

Fig.4. The dielectric constant as a function of frequency recorded at room temperature for PMMA/CCTO composite.

The room temperature dielectric loss (D) recorded as a function of frequency for PMMA, and PMMA/CCTO composites are shown in Fig.5. The dielectric loss of PMMA polymer has been increased as the CCTO content increased, but decreases over the frequency. Around 10kHz, it seen that (marked) there is a sudden drop in the loss value. Since the addition of fillers induces structural changes in the polymer [10], sudden drop in the dielectric loss was observed for all the composites under study. At low frequency, the dielectric loss is considerably higher, due to the interfacial polarization/MWS effect [10]. The dielectric loss for all the composites lies below 0.1 for the entire frequency under investigation. The dielectric loss obtained for pure PMMA is almost independent of the frequency (100Hz to 100MHz). In this work, remarkably low dielectric loss was achieved for the PMMA+CCTO composites. The loss value obtained @100 Hz for PMMA+CCTO-40 Vol % composite is around 0.094 and it has decreased to 0.0094 @10MHz.

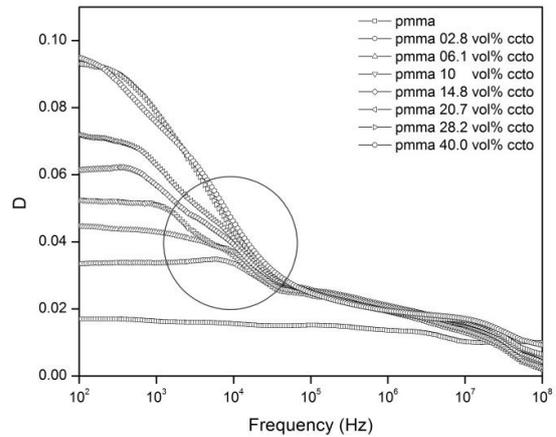

Fig.5. The dielectric loss as a function of frequency recorded at room temperature for PMMA/CCTO composite.

The possibility of using the PMMA+CCTO composite at high frequency capacitors application could be explored. Overall, the dielectric constant of the composite increases as the ceramic content increases, and much lower than that of the ceramics. It is to be noted that for the fabrication of PMMA-CCTO composite, unsintered CCTO powder has been employed. It is well known that the dielectric constant for the sintered ceramic is very high compared to that of powder one, we find that the dielectric constant obtained for the composites is much lower than that of the sintered ceramic. It has been reported that the interphase region in polymer-ceramic composites play an important role and it can be modified by the addition of metal particulates, thereby increasing the conductivity of the polymer, resulting in the high dielectric constant of the composite [2,17]. Keeping this in mind, efforts are on for the fabrication of PMMA+M+CCTO (where M=Ni,Cu and Al) three phase composites systems with improved dielectric properties suitable for capacitors in embedded passive devices.

## CONCLUSIONS

Composites with varying concentrations of CCTO (2.8 to 40 %) by vol in Poly(methyl methacrylate) (PMMA) polymer matrix by melt mixing and hot pressing process were fabricated. The dielectric constant of PMMA increased with increase in CCTO content ceramic. The PMMA+CCTO-40Vol % is higher than that of pure PMMA. The low frequency relaxation is attributed to the space charge polarization/MWS effect. PMMA+CCTO composites exhibited remarkably low dielectric loss, which can be exploited for the high frequency capacitors application.

## ACKNOWLEDGEMENT

The management of Central Power Research Institute is acknowledged for the financial support (CPRI Project No. R-DMD-01/1415).